%% file: paper.tex
\documentclass[copyright]{eptcs}
\providecommand{\event}{DSL'11} % Name of the event you are submitting to
\usepackage[utf8]{inputenc}
\usepackage[T1]{fontenc}
\usepackage{graphicx}
\usepackage[compact]{titlesec}
\usepackage{listings}
\usepackage{multicol}
\usepackage{subfigure}

\DeclareMathAlphabet{\mathtt}{OT1}{txtt}{m}{n}
\SetMathAlphabet{\mathtt}{bold}{OT1}{txtt}{b}{n}

\lstdefinelanguage{Scala}%
{morekeywords={abstract,%
  case,catch,char,class,%
  def,else,extends,final,for,%
  if,import,implicit,%
  match,module,%
  new,null,%
  object,override,%
  package,private,protected,public,%
  for,public,return,super,%
  this,throw,trait,try,type,%
  val,var,%
  while,with,%
  yield%
  },%
  sensitive,%
  morecomment=[l]//,%
  morecomment=[s]{/*}{*/},%
  morestring=[b]",%
  morestring=[b]',%
  showstringspaces=false%
}[keywords,comments,strings]%

\lstnewenvironment{slisting}{\lstset{language=Scala}}{}
\lstnewenvironment{listing}{\lstset{language=Scala}}{}

\newcommand{\scode}[1]{\lstinline[language=Scala,columns=fixed,basicstyle=\ttfamily,keywordstyle=\ttfamily]|#1|}

\newcommand{\code}[1]{\scode{#1}}

\newcommand{\comment}[1]{}

\title{Building-Blocks for Performance Oriented DSLs}
\author{Tiark Rompf$^\ast$ \quad Arvind K. Sujeeth$^\dagger$  \quad HyoukJoong Lee$^\dagger$  
\quad Kevin J. Brown$^\dagger$  \quad Hassan Chafi$^\dagger$ \\
Martin Odersky$^\ast$ \quad Kunle Olukotun$^\dagger$ 
\institute{$^\ast$ EPFL \quad $^\dagger$ Stanford University}
\email{\{first.last\}@epfl.ch \quad \{asujeeth,hyouklee,kjbrown,hchafi,kunle\}@stanford.edu}
}
\def\titlerunning{Building-Blocks for Performance Oriented DSLs}
\def\authorrunning{T. Rompf, A. Sujeeth, H. Lee, K. Brown, H. Chafi, M. Odersky, K. Olukotun}
\def\copyrightholders{Rompf, Sujeeth, Lee, Brown, Chafi, Odersky, Olukotun}
\begin{document}
\maketitle

\begin{abstract}
Domain-specific languages raise the level of abstraction in software development.
While it is evident that programmers can more easily reason about very high-level
programs, the same holds for compilers only if the compiler has an accurate model
of the application domain and the underlying target platform. 
Since mapping high-level, general-purpose languages to modern, heterogeneous hardware 
is becoming increasingly difficult, DSLs are an attractive way to
capitalize on improved hardware performance, precisely by making the compiler 
reason on a higher level. Implementing efficient DSL compilers is a daunting task
however, and support for building performance-oriented DSLs is urgently needed. 
To this end, we present the Delite Framework, an extensible toolkit that drastically 
simplifies building embedded DSLs and compiling DSL programs for execution on 
heterogeneous hardware. We discuss several building blocks in some detail and
present experimental results for the OptiML machine-learning DSL implemented on
top of Delite.
\end{abstract}

\input{intro}

\input{lms}

\input{delite}

\input{main}

\input{optiml}

\input{related}

\input{conclusions}

%\nocite{*}
%\bibliographystyle{eptcs}
%\bibliography{generic}

\bibliographystyle{eptcs}
\bibliography{ppl.bib}

\end{document}

%% file: intro.tex
\section{Introduction}

Generic high-level programming languages are no longer compiled efficiently to modern hardware,
which is increasingly parallel and often consists of heterogeneous processing
elements, i.e.\ multiple CPUs and GPUs \cite{hudak96building,scala,intelArBB,sheardTemplateHaskell}.
%
%\note{pact}Current microprocessor trends focus on larger numbers of sim- pler cores [22, 35] and include increasingly heterogeneous pro- cessing elements, such as SIMD units or a GPU [2, 41]. These heterogeneous architectures are continuing to provide increases in maximum achievable performance, but unfortunately programming these devices to reach these performance levels is not straightfor- ward. Each heterogeneous element has its own performance char- acteristics and pitfalls, and usually comes with its own program- ming model. Therefore when targeting such architectures, the pro- grammer must have a deep understanding of all the different hard- ware components and programming models, as well as understand how to make all them all work together. Even with this understand- ing, the best way of dividing work across the hardware is often unpredictable and affected by application variables such as dataset size, making it nontrivial to realize maximal performance improve- ment.\note{end pact}
%
%Programming for emerging hardware devices is becoming increasingly difficult as architectures 
%become more parallel and heterogeneous \cite{??}.  General-purpose programming languages 
%typically either do not support these exotic \note{ becoming mainstream! } architectures or require the programmer to 
%manage all the low-level details.  
Without adequate support from their favorite high-level language, programmers wishing to fully exploit
today's hardware have no choice but to utilize low-level, hardware-specific programming models, 
such as Pthreads for multi-core, CUDA or OpenCL for GPUs, and MPI for clusters.
Not only does a programmer have to understand how to use all of these disparate low-level
programming models individually, he or she must understand how best to combine them for a given 
application.  
These decisions are in general not straightforward and often depend on variables such as 
dataset size. Furthermore, having to make these choices and commit to specific programming
models severely limits the portability and maintainability of the application. 
Overall, the result is that developing performance-oriented applications greatly diminishes
programmer productivity.

%Given the limitations of current general-purpose languages in these 
%situations, domain-specific languages (DSLs) provide an attractive alternative.  

%DSLs increase programmer productivity by offering extremely high-level abstractions.
%Programmers can more easily reason about these.
%
%\textbf{Performance-oriented DSLs:} make the \emph{compiler} more productive (produce
%better code) by reasoning at a higher level.
Domain-specific languages (DSLs) provide an attractive alternative.
DSLs have a long history of increasing programmer productivity by 
providing extremely high-level, in a sense ``ideal'', abstractions tailored to a 
particular domain. 
Performance-oriented DSLs strive to also make the \emph{compiler} more 
productive (producing better code) by enabling it to reason on a higher level
as well. 
%
%While it is evident that programmers can more easily reason about high-level
%abstractions, compilers must be fitted with an intimate understanding of the domain
%operations and data structures.
%
Interestingly, 
while productivity and performance are often at odds in general-purpose languages, the 
higher level of abstraction provided by DSLs and willingness to sacrifice generality makes 
it feasible for a DSL compiler and runtime system to generate high performance code, 
including code targeting parallel heterogeneous architectures, from high-level, 
single-source application code \cite{ppopp11delite}.

%These performance-oriented DSLs are successful largely due to making the \emph{compiler} more 
%productive (producing better code) by enabling higher level reasoning across domain 
%operations, data structures, and other constructs.  This is in stark contrast to general-
Fitting a compiler with intimate knowledge about domain operations, data structures, and 
other constructs is in stark contrast to general-purpose 
languages which focus on primitives for abstraction and composition, such that programmers
can build large systems from few and relatively simple but versatile parts.
Consequently, general-purpose compilers do not understand the semantics of the 
complex operations performed within an application. Reasoning across domain constructs, however,
enables more powerful and aggressive optimizations that are infeasible otherwise.
General-purpose languages impose very few restrictions on programmers which 
in turn requires the compiler to perform non-trivial analyses to prove that  
%classic 
optimizations are safe.  Unfortunately, safety of optimizations often cannot be 
determined and therefore the compiler must be conservative and not apply the optimization
to guarantee correctness of the generated code. This 
ultimately leads to programmers having to perform a large amount of obscure performance 
debugging that can require reverse-engineering the compiler's algorithms to 
determine why a certain piece of code runs slowly. Slight changes in the input program
can have large effects on performance so performance debugging effort is not a 
one-time effort but a continuous one.
The situation is complicated further if multiple layers of compilation
(and possibly profile-driven re-compilation)
%(e.g.\ ahead of time and profile-driven just in time) 
are involved as is the case for 
managed runtimes such as the Java Virtual Machine (JVM).
Performance-oriented DSLs can take the 
opposite approach to the problem, namely restrict the programmer from writing code that 
would prevent the compiler from generating an efficient implementation.  The compiler is 
then able to perform very aggressive optimizations with much simpler or even without safety
analyses, providing the programmer with efficient code for significantly less effort. 

While the promise of creating performance-oriented DSLs capable of targeting emerging 
heterogeneous architectures is high, building these DSLs remains a significant challenge.  
The first and most obvious challenge is constructing a new language and compiler from scratch. 
While much of the novelty and usefulness of DSLs lies in the domain-specific aspects of the 
language and compiler, a non-trivial portion of the work for the developer lies in 
re-implementing more general-purpose features (e.g., parsing, type checking, performing generic 
optimizations, etc.).  Furthermore, the DSL developer must be not only a domain expert, but 
also an expert in parallelism and architecture to properly optimize for modern 
heterogeneous hardware.  Rather than starting afresh for each new DSL, a DSL developer should 
be able to construct a new language from building blocks that are common across a variety 
of DSLs and focus on adding domain-specific constructs on top of these general building 
blocks.  One way to achieve this goal is to embed new DSLs within a highly expressive 
general-purpose host language. %\comment{\note{transition}}

In essence, we want to obtain the flexibility and performance achievable 
with external (stand-alone) DSLs while maintaining the modest effort required in creating 
purely-embedded internal (library-based) DSLs, a notion we have termed 
\emph{language virtualization} \cite{hassan10virtualization}.
From this principle we have developed \emph{lightweight modular 
staging} (LMS) \cite{rompf10lms} as a means of building new embedded DSLs in Scala \cite{scala} and 
creating optimizing domain-specific compilers at the library level.  
LMS is a multi-stage programming \cite{Walid:1999:MPT} approach, i.e.\ a disciplined form
of runtime code generation. Unlike dedicated multi-stage languages such as MetaML \cite{DBLP:journals/tcs/TahaS00}, 
LMS does not employ syntactic annotations to designate staged expressions but instead
relies on type signatures, very similar to and inspired by \emph{finally tagless} \cite{DBLP:conf/aplas/CaretteKS07} or 
\emph{polymorphic} embedding of DSLs \cite{hoffer08polymorphic}. 
Furthermore, LMS uses overloaded operators to combine staged code fragments in a 
semantic way unlike quasi-quotation approaches that are merely syntactic expanders.
Overloading of operators thus provides a natural and principled interface for generic and 
domain-specific optimizations. LMS also comes with strong well-formedness and typing guarantees, 
most of them inherited from the \emph{finally tagless} \cite{DBLP:conf/aplas/CaretteKS07}
embedding of typed object languages into types metalanguages. Any well-typed program
generator will produce well-formed and well-typed code, unless maliciously subverted
through explicit type casts, Java reflection, or other inherently unsafe mechanisms.
On top of LMS we have developed Delite, which provides compilation and runtime 
support for execution on heterogeneous targets (i.e., multi-core CPU and GPU) to the DSL 
developer.

%The paper introduces "Lightweight Modular Staging" (LMS), a technique for expressing multi-stage programming. Staged programming allows control over the time when a piece of code is executed. In particular, code can be executed at compile time already, thereby allowing a disciplined form of program generation. Most programming languages that support staging require explicit source code annotations that delay or force the execution of a marked piece of code. LMS is different in that it does not have annotations on the term level: instead, simply changing the type signatures of a program fragment is sufficient to lift that program to a staged version. Additional optimizations can then be implemented separately (using Scala's "trait"s), without touching the original code. LMS is particularly attractive for implementing embedded domain-specific languages (DSLs), because the staging features can be used to perform code optimizations in the background and the lightweight syntax keeps the surface syntax clean. This is confirmed by the fact that the Stanford Pervasive Parallelism Lab is already using LMS successfully to build domain-specific languages. The paper is a good example how modern language technology can help to solve long-standing and difficult problems such as staging: with Scala's general-purpose features, the staging system as described in this paper can be implemented as a library, without need for further language extensions.

The rest of this paper is organized as follows. In Section~\ref{sec:delite} we 
discuss how LMS aids in creating a new domain-specific language and optimizing compiler 
through the use of extensible, composable building blocks.  We then discuss how Delite 
extends LMS to provide structured parallel patterns and code generation support to 
make creating a performance-oriented DSL targeting heterogeneous parallel architectures 
only incrementally more difficult than targeting a traditional uniprocessor.  In Section 
\ref{sec:main} we illustrate how we can use DSLs and the LMS/Delite framework to perform 
various kinds of optimizations. Section~\ref{sec:optiml} presents a case study of the
OptiML machine learning DSL, including performance evaluation.
We then discuss related work in Section \ref{sec:related} and conclude in
Section \ref{sec:conclusions}.

%\textbf{Problems compiling general purpose language:}

%Very different views of the program. Programmer: "do task X". Hl-Language: Objects,
%classes, higher-order functions. Hardware: Memory, parallelism, etc.

%Lowest level: Hardware is too complex and diverse (heterogeneous, etc).

%Middle level: programming languages are about abstraction and composition.
%Build large systems out of small parts.

%Optimizing compilers need to do analysis that ``sees through'' abstraction mechanisms (hard to do).
%Most of the time they don't see the forest for the trees.

%General-purpose compilers need to be conservative: err on the side of correctness. Do not apply optimizations
%unless proven to be safe. Compile any legal program to executable code, even 
%if optimizations cannot be applied (so no guaranteed optimizations usually).
%Also: users don't get feedback on what got optimized, what didn't and why.

%Consequence for writing performance-critical code: can't use very high-level languages but need 
%hardware-specific programming models (CUDA, pthreads, ...).
%Tough times for programmers: need to go more and more
%low-level for performance, and not only once but anew for each programming model. 

%Performance-oriented DSLs: teach compiler about domain. Take liberties to be more restrictive:
%sometime we want to reject programs that cannot be optimized in required ways.

%\textbf{Building performance-oriented DSLs is really hard.} First and foremost because one
%has to start from scratch. We need building blocks out of which we can construct them.

%Delite/LMS to the rescue...

%% file: lms.tex
\section{An End-to-End System for Embedded Parallel DSLs}
\label{sec:delite}

This section gives an overview of our approach to developing and executing
embedded DSLs in parallel and on heterogeneous devices. We have built reusable
infrastructure to alleviate the burden of building a high performance
DSL. We embed our DSLs in Scala using Lightweight Modular
Staging~\cite{rompf10lms}, and provide a common intermediate representation (IR) and basic
facilities for optimization and code generation. On top of this layer, we have developed
Delite, a toolbox for creating parallel DSLs. Delite is structured into a
\emph{framework} and a \emph{runtime} component.  The framework provides
primitives for parallel operations such as \code{map} or \code{reduce} that DSL
authors can use to define higher-level operations. Once a DSL author uses
Delite operations, Delite handles code generating to multiple platforms (e.g.
Scala and CUDA), and handles difficult but common issues such as device
communication and synchronization. These capabilities are enabled by exploiting
the domain-specific knowledge and restricted semantics of the DSL compiler.

\subsection{Building an IR using Lightweight Modular Staging}
\label{subsec:lms}

%\note{TODO: start more high-level, summary...}
On the surface, DSLs implemented on top of Delite appear very similar to
purely-embedded (i.e.\ library only) Scala-based DSLs.
However, a key aspect of LMS and hence Delite is that DSLs are split in two parts,
\emph{interface} and \emph{implementation}. Both parts can be assembled from components in
the form of Scala traits.
DSL programs are written in terms of
the DSL interface, agnostic of the implementation. Part of each DSL interface is an abstract
type constructor \code{Rep[_]} that is used to wrap types in DSL programs.
For example, DSL programs use \code{Rep[Int]} wherever a regular program would
use \code{Int}. The DSL operations defined in the DSL interface (most of them are
abstract methods) are all expressed in terms of \code{Rep} types. 

The DSL \emph{implementation} provides a concrete instantiation of \code{Rep} as
expression trees (or graphs). The DSL operations left abstract in the interface are
implemented to create an expression representation of the operation.
Thus, as a result of executing the DSL program, we obtain an analyzable
representation of the very DSL program which we will refer to as IR
(intermediate representation).

%On the surface, DSLs implemented on top of Delite are very similar to
%regular purely-embedded (i.e.\ library only) Scala-based DSLs. 
To substantiate the description, let us consider an example step by step.
A simple (and rather pointless) program that calculates the average of 
100 random numbers, written in a prototypical DSL \code{MyDSL} that includes numeric
vectors and basic console IO could look like this:
\begin{slisting}
  object HelloWorldRunner extends MyDSLApplicationRunner with HelloWorld
  trait HelloWorld extends MyDSLApplication {
    def main() = {
      val v = Vector.rand(100)
      println("today's lucky number is: ")
      println(v.avg)
    }
  }
\end{slisting}
Programs in our sample DSL live within traits that inherit from \code{MyDSLApplication},
with method \code{main} as the entry point. In Scala, traits are similar to
classes but can participate in mixin-composition \cite{scala}. 
Scala's mixin composition is a restricted form of multiple inheritance that resolves
super calls according to an inheritance-preserving linearization of all the receiver's
base traits and classes.
%Using mixin composition, the new member definitions of a trait (i.e.\ the delta in relationship to the supertrait or
%-class) can be reused

\code{MyDSLApplication} is a trait provided by the DSL that defines the DSL interface.
In addition to the actual DSL program, there is a singleton object that inherits
from \code{MyDSLApplicationRunner} and mixes in the trait that contains the program.
As the name implies, this object will be responsible for directing the
staged execution of the DSL application.

Here is the definition of \code{MyDSL}'s components encountered so far:
\begin{slisting}
  trait MyDSLApplication extends DeliteApplication with MyDSL
  trait MyDSLApplicationRunner extends DeliteApplicationRunner with MyDSLExp

  trait MyDSL extends ScalaOpsPkg with VectorOps
  trait MyDSLExp extends ScalaOpsPkgExp with VectorOpsExp with MyDSL
\end{slisting}

\code{MyDSLApplicationRunner} inherits
the mechanics for invoking code generation from DeliteApplication. We discuss how
Delite provides these facilities in section~\ref{subsec:delite}.
We observe a structural split in the inheritance hierarchy that is rather fundamental:
\code{MyDSL} defines the DSL \emph{interface}, \code{MyDSLExp} the \emph{implementation}.
A DSL program is written with respect to the interface but it knows nothing
about the implementation. The main reason for this separation is safety.
If a DSL program could observe its own structure, optimizing rewrites
that maintain semantic but not structural equality of DSL expressions
could no longer be applied safely.\footnote{In fact, this is the main 
reason why MSP languages do not allow inspection of staged code at all \cite{DBLP:conf/pepm/Taha00}.}
Our sample DSL includes a set of common Scala operations that are provided 
by the core LMS library as trait \code{ScalaOpsPkg}. These operations
include conditionals, loops, variables and also \code{println}.
On top of this set of generic things that are inherited from Scala,
the DSL contains vectors and associated operations. 
The corresponding interface is defined as follows:

%\begin{slisting}
%  trait DeliteApplication {
%    def main(): Unit
%  }
%  trait DeliteApplicationRunner extends DeliteOpsExp with DeliteApplication {
%    val generators: List[GenericCodegen{ val IR: DeliteApplication.this.type }] // list of code generators
%    def main(args: Array[String]): Unit = {
%      // stage the main() function and generate code
%    }
%  }
%\end{slisting}

\begin{slisting}
  trait VectorOps extends Base {
    abstract class Vector[T]                                  // placeholder ("phantom") type
    object Vector {
      def rand(n: Rep[Int]) = vector_rand(n)                  // invoked as: Vector.rand(n)
    }
    def vector_rand(n: Rep[Int]): Rep[Vector[Double]]
    def infix_length[T](v: Rep[Vector[T]]): Rep[Int]          // invoked as: v.length
    def infix_sum[T:Numeric](v: Rep[Vector[T]]): Rep[T]       // invoked as: v.sum
    def infix_avg[T:Numeric](v: Rep[Vector[T]]): Rep[T]       // invoked as: v.avg
    ...
  }
\end{slisting}

There is an abstract class \code{Vector[T]} for vectors with element type \code{T}. The notation
\code{T:Numeric} means that \code{T} may only range over numeric types such as \code{Int} or \code{Double}.
Operations on vectors are not declared as instance methods of \code{Vector[T]} but as external functions over
values of type \code{Rep[Vector[T]]}.
Likewise, when referring to primitive values, \code{Rep[Int]}
is used instead of \code{Int}. This wrapping of types is the core LMS abstraction. 
An expression of type \code{Rep[T]} denotes an expression that \emph{represents} the computation of
a value of type \code{T}, i.e.\ will produce a value of type \code{T} in the next computation stage, 
when the generated code is executed. The root of the interface hierarchy, trait \code{Base} from 
the core LMS library, defines \code{Rep} as an abstract type constructor:
\begin{slisting}
  trait Base {
    type Rep[T]
  }
\end{slisting}

An instance of a \code{Rep[T]} can be constructed by a DSL factory method directly, or lifted from
a concrete instance using an implicit conversion. Concrete instances (such as an integer) are constants
in the IR, since they already existed before being passed to a staged DSL method.

The corresponding implementation counterpart, trait \code{BaseExp}, defines the concrete
intermediate representation (IR) that is used by the core LMS library and consequently Delite:
\begin{slisting}
  trait BaseExp extends Base with Expressions {
    type Rep[T] = Exp[T]
  }
  trait Expressions {
    // expressions (atomic)
    abstract class Exp[T]
    case class Const[T](x: T) extends Exp[T]
    case class Sym[T](n: Int) extends Exp[T]

    // definitions (composite, subclasses provided by other traits)
    abstract class Def[T]

    // bind definitions to symbols automatically
    implicit def toAtom[T](d: Def[T]): Exp[T] = ...
  }
\end{slisting}

In contrast to other approaches based on abstract type constructors \cite{hoffer08polymorphic} we do not use
multiple DSL representations but a single, extensible one. 
Subtraits of \code{BaseExp} are free to add subclasses of \code{Def}. This model allows generic
optimizers to view the IR in terms of its base nodes (\code{Exp},\code{Def}), while DSL subclasses
can extend these nodes with richer semantic information and use them at a higher level.

Returning to our sample DSL, this is the definition of \code{VectorOpsExp}, the
implementation counterpart to the interface defined above in \code{VectorOps}:

\begin{slisting}
  trait VectorOpsExp extends DeliteOpsExp with VectorOps {
    case class VectorRand[T](n: Exp[Int]) extends Def[Vector[Double]]
    case class VectorLength[T](v: Exp[Vector[T]]) extends Def[Int]

    case class VectorSum[T:Numeric](v: Exp[Vector[T]]) extends DeliteOpLoop[Exp[T]] {
      val range = v.length
      val body = DeliteReduceElem[T](v)(_ + _) // scalar addition (impl not shown)
    }

    def vector_rand(n: Rep[Int]) = VectorRand(n)
    def infix_length[T](v: Rep[Vector[T]]) = VectorLength(v)
    def infix_sum[T:Numeric](v: Rep[Vector[T]]) = VectorSum(v)
    def infix_avg[T:Numeric](v: Rep[Vector[T]]) = v.sum / v.length
    ...
  }
\end{slisting}
The constructor \code{rand} and the function \code{length} are implemented as
new plain IR nodes (extending \code{Def}). Operation \code{avg} is implemented directly in terms of \code{sum}
and \code{length} whereas \code{sum} is implemented as a \code{DeliteOpLoop} 
with a \code{DeliteReduceElem} body. 
These special classes of structured IR nodes are provided by the Delite framework
and are inherited via \code{DeliteOpsExp}.

A closing note on type safety is in order. Since all DSL operations are expressed as typed Scala methods, 
the Scala type system ensures that DSL operations are used in a type-safe way in DSL programs.
Trying to invoke \code{v.sum} where \code{v} is is a vector of Strings (type \code{Rep[Vector[String]]}) would
be a compile-time type error, since \code{String} is not a numeric type (there is no instance of \code{Numeric} for
type \code{String}). On the implementation side, the Scala type system makes sure that \code{infix\_sum} creates a typed IR
node that corresponds to the declared return type \code{Rep[Vector[T]] = Exp[Vector[T]]}. Thus, the types of the
IR nodes correspond to the types in the DSL program. 
Transformations on the IR need to ensure that the types of IR nodes are preserved. Again, the typed
embedding helps because the Scala type system ensures that the result of a transformation conforms 
to the given method signatures (i.e.\ by implying that a transformation maps \code{Exp[T]} to 
\code{Exp[T]} for all types \code{T}).

\subsection{Code Generation}
\label{subsec:codegen}

The LMS framework provides a code generation infrastructure that includes a program scheduler
and a set of base code generators. The program scheduler uses the data
and control dependencies encoded by IR nodes to determine the sequence of nodes that should be
generated to produce the result of a block. 
%The scheduler performs code motion optimizations,
%including hoisting computation out of loops and pushing it into conditionals when possible. Since programs
%are scheduled only by their dependencies, dead code is also eliminated at this stage. 
After the scheduler has determined a schedule, it invokes
the code generator on each node in turn. There is one \emph{code generator} object
per target platform (e.g. Scala, CUDA, C++) that mixes together traits that describe how to generate
platform-specific code for each IR node. This organization makes it easy for DSL authors to modularly extend the base code generators;
they only have to define additional traits to be mixed in with the base generator. 

To be more concrete, consider the following definition of GenericCodegen, which all code generators
extend:

\begin{slisting}
  trait GenericCodegen extends Scheduling {
    val IR: Expressions
    import IR._
    def emitBlock(y: Exp[Any])(implicit stream: PrintWriter): Unit = {
      val deflist = buildScheduleForResult(y)
      for (TP(sym, rhs) <- deflist) {
        emitNode(sym, rhs)
      }
    }
    def emitNode(sym: Sym[Any], rhs: Def[Any])(implicit stream: PrintWriter): Unit = {
      throw new GenerationFailedException("don't know how to generate code for: " + rhs)
    }
  }
\end{slisting}

A more sophisticated version extends GenericCodegen to maintain a scope, which allows emitBlock to be called
in a nested fashion. The code generator is not part of the program object that contains the DSL IR definitions. 
This separation allows multiple code generators to be invoked on a particular program, but requires a mechanism
to inject the definition of \code{Rep} and the IR nodes into the code generator object. We handle this using
the \code{IR} value, which represents a path-dependent type. The IR value is instantiated when the code generator
is created. Given the VectorOpsExp definition from section~\ref{subsec:lms}, we could inject the IR as follows:

\begin{slisting}
  val myCodeGenerator = new GenericCodegen { val IR: VectorOpsExp.this.type = VectorOpsExp.this }
\end{slisting}

Continuing the VectorOps example, we can extend the LMS base Scala generator to generate field lookups on a Vector
instance:

\begin{slisting}
  trait ScalaGenVectorOps extends ScalaGenBase {
    val IR: VectorOpsExp
    import IR._

    override def emitNode(sym: Sym[Any], rhs: Def[Any])(implicit stream: PrintWriter) = rhs match {
      case VectorLength(x)    => emitValDef(sym, quote(x) + ".length")
      case _ => super.emitNode(sym, rhs)
    }
  }
\end{slisting}
 
Now we can compose this together with the Scala code generators provided by LMS:

\begin{slisting}
  trait MyDSLGen extends ScalaCodeGenPkg with ScalaGenVectorOps {
    val IR: MyDSLOpsExp
  }
\end{slisting}

Therefore, DSL designers only have to add code generators for their own
domain-specific types. They inherit the common functionality of scheduling and
callbacks to the generation methods, and can also build on top of code
generator traits that have already been defined. In many cases, though, DSL authors do not
have to write code generators at all; the next section describes how Delite 
takes over this responsibility for most operations.

%% file: delite.tex
\subsection{The Delite Compiler Framework and Runtime} 
\label{subsec:delite}
%On top of the LMS framework that provides the facility to construct IR nodes for DSL
%operations, the Delite Compiler Framework that significantly
%reduces the burden of building a new DSL. One of the most challenging tasks
%when building a high performance parallel DSL is generating efficient parallel
%code for heterogeneous targets since it requires the expertise in parallel
%programming and hardware. 

On top of the LMS framework that provides the basic means to construct IR nodes
for DSL operations, the Delite Compiler Framework provides high-level
representations of execution patterns through \code{DeliteOp} IR, which
includes a set of common parallel execution patterns (e.g. map, zipWith,
reduce).

\code{DeliteOp} extends \code{Def}, and DSL operations may extend one of the
\code{DeliteOps} that best describes the operation.  For example, since 
\code{VectorSum} has the semantics of iterating over the elements of the input
Vector and adding them to reduce to a single value, it can be
implemented by extending \code{DeliteOpLoop} with a reduction operation as its
body. This significantly reduces the amount of work in implementing a
DSL operation since the DSL developers only need to specify the
necessary fields of the \code{DeliteOp} (\code{range} and \code{body} in the case of
\code{DeliteOpLoop}) instead of fully implementing the operation.  

\code{DeliteOpLoop}s are intended as parallel for-loops. Given an
integer index range, the runtime guarantees to execute the loop body exactly once 
for each index but does not guarantee any execution order. 
Mutating global state from within a loop is only safe at disjoint indexes. 
There are specialized constructs to define loop bodies for map and reduce patterns (\code{DeliteCollectElem}, \code{DeliteReduceElem})
that transform a collection of elements point-wise or perform aggregation. 
An optional predicate can be added to perform filter-style operations, i.e.\ select or aggregate only those
elements for which the predicate is true. All loop constructs can be fused into 
\code{DeliteOpLoops} that do several operations at once.

Given the relaxed ordering guarantees, the framework can automatically generate 
efficient parallel code for \code{DeliteOps}, targeting heterogeneous parallel hardware.  
Therefore, DSL developers can easily implement parallel DSL operations by
extending one of the parallel \code{DeliteOps}, and only focus on the
language design without knowing the low-level details of the target
hardware.  Below is the code snippet for the \code{DeliteOpLoop} Scala target generator.
The code is simplified to focus on the case where the \code{body} of \code{DeliteOpLoop} is of type \code{DeliteReduceElem}.
The generated kernel creates an object of type \code{generated.scala.DeliteOpLoop} that defines methods for
creating the output object (\code{alloc}) and processing the element of the input collection (\code{process, combine}).
Those methods will be called by the runtime during execution.

%      // generate process method for each element	  
%      stream.println("def process(__act: " + actType + ", " + quotearg(op.v) + "): Unit = {")
%      < ... > 
%      stream.println("}")

\begin{slisting}
trait ScalaGenDeliteOps extends BaseGenDeliteOps {
  import IR._
  override def emitNode(sym: Sym[Any], rhs: Def[Any])(implicit stream: PrintWriter) = rhs match {
    case DeliteOpSingleTask[_] =>  ... 
    case op@DeliteOpLoop[_] =>  
      stream.println("val " + kernelName + " = new generated.scala.DeliteOpLoop[" + actType + "] {")
      stream.println("def size = " + quote(op.range))   // input size
      stream.println("def alloc: " + actType + " = {")  // output allocation
      stream.println("val __act = new " + actType)      // activation record (environment)
      (symList zip op.body) foreach {
        case (sym, elem: DeliteCollectElem[_,_]) => ...
        case (sym, elem: DeliteReduceElem[_]) => 
          stream.println("__act." + quote(sym) + " = " + quote(elem.zero))
      }
      stream.println("__act")
      stream.println("}")
      ... 
      // generate reduction method     
      stream.println("def combine(__act: " + actType + ", rhs: " + actType + "): Unit = {")
      (symList zip op.body) foreach {
        case (sym, elem: DeliteCollectElem[_,_]) => ...
        case (sym, elem: DeliteReduceElem[_]) =>
          stream.println("val " + quote(elem.rV._1) + " = " + "__act." + quote(sym))
          stream.println("val " + quote(elem.rV._2) + " = " + "rhs." + quote(sym))
          emitBlock(elem.rFunc)
          stream.println("__act." + quote(sym) + " = " + quote(getBlockResult(elem.rFunc)))
      }
      stream.println("}")

    case _ => super.emitNode(sym, rhs)
  }
}
\end{slisting}

It is important to note that \code{emitNode} is the very last step in the program generation process and 
until then, everything is well-typed.
The use of strings to assemble an (untyped) source code representation of the generated program 
is unavoidable if the final target is source code and does not limit the overall safety in any way.
Strings are used ``write-only'' for output, they are never stored or manipulated otherwise.

%To enable those features, the Delite Compiler Framework presents a multi-view
%notion of IR. Each IR node can be viewed in three different perspectives, each
%serving for different types of optimizations and code generations. Figure X

%In addition, the framework optimizes the IR with traditional compiler
%optimizations (e.g., common subexpression elimination), and also provides an
%interface for the DSL developers to add domain-specific optimizations (e.g.,
%linear algebra simplifications) through pattern matching.
%
%To enable those features, the Delite Compiler Framework presents a multi-view
%notion of IR. Each IR node can be viewed in three different perspectives, each
%serving for different types of optimizations and code generations. Figure X
%describes the structure of this notion in the framework. A generic IR node
%encodes a symbol with the corresponding definition, and this view allows for
%the traditional compiler optimizations such as CSE and code motion. The
%parallel IR view is to provide a set of parallel execution patterns for
%implementing the DSL operations and to automatically generate parallel code for
%heterogeneous targets. A Domain-specific (DS) IR node encodes the
%domain-information for domain-specific optimizations. 

The Delite Compiler Framework currently supports Scala, C++, and CUDA targets.
The framework provides code generators for each target in addition to a main
generator (\emph{Delite generator}) that controls them. The \emph{Delite generator} iterates
over the list of available target generators to emit the target-specific
kernels. By generating multiple target implementations of the kernels and
deferring the decision of which one to use, the framework provides the runtime with
enough flexibility in scheduling the kernels based on dynamic information such as 
resource availability and input size. In addition to the kernels, the
Delite generator also generates the \emph{Delite Execution Graph} (DEG) of the
application. The DEG is a high-level representation of the program that
encodes all necessary information for its execution, including
the list of inputs, outputs, and interdependencies of all kernels.  
%\note{todo: relate to traits DeliteApplication and DeliteApplicationRunner from above}

After all the kernels are generated, the Delite Runtime starts analyzing the
DEG and emits execution plans for each target hardware, taking the machine
status (e.g.\ number of available CPUs and GPUs) into account. An execution plan
consists of a series of kernel calls and necessary synchronizations with the
kernels in other execution plans. When two targets share data but have separate
address spaces, the runtime scheduler inserts data transfer operations whenever
necessary. Since the DEG encodes all the dependencies of the kernels, the
scheduler can statically determine the only places where data
transfers are needed through liveness analysis. The scheduler also uses a
heuristic based on a clustering algorithm \cite{sinnen2007task}
to minimize the communication between targets. 
When the scheduling decisions are made and execution plans are
emitted, the runtime finally invokes the target compilers to generate
executables for each target.

%% file: main.tex
\section{Performance Building-Blocks}
\label{sec:main}

In this section we take a closer look at some aspects of the approach outlined in
Section~\ref{sec:delite} that
work together in new or interesting ways to support the development of performance-oriented DSLs.
Broadly, the important aspects fall into two categories:

\textbf{Artifacts:} Reusable pieces of software provided by the Delite and LMS framework. These
include `standard' compiler optimizations, a generic facility for pattern rewrites,
\emph{DeliteOps} and other IR definitions for generic functionality and easy parallelization,
loop fusion (see Section~\ref{subsec:fusion}), and a scheduler that tracks data and control
dependencies.

\textbf{Concepts:} Design pattern and practices that aid the development of DSLs. Among these
are the use of multi-stage programming, i.e.\ viewing DSL programs as program generators and
using abstraction in the generator instead of abstraction in the DSL implementation.
Another pattern is the overall organization into independent component, such that 
IR nodes, optimizations and code generators can be composed in a modular way.

Discussing all of these items in sufficient depth would be beyond the scope of a single
paper. Therefore the following sections are to be understood as a selection.

\subsection{Regular Compiler Optimizations}

Many classic compiler optimizations can be applied to the IR generated from DSL
programs in a straightforward way. Among the generic optimizations that are applied
by default are common subexpression elimination, dead code elimination, constant
folding and code motion. Due to the structure of the IR, these optimizations all
operate in an essentially global way, at the level of domain operations.
An important difference to regular general-purpose compilers is that
IR nodes carry information about effects they incur (see below). 
This allows to use quite
precise dependency tracking that provides the code generator with a lot of
freedom to group and rearrange operations. Consequently, optimizations like
common subexpression elimination and dead code elimination will
easily remove complex DSL operations that contain internal control-flow and
may span many lines of source code. The same holds for code motion.
Consider the following user-written code:
\begin{listing}
  v1 map { x =>
    val s = sum(v2.length) { i => v2(i) }
    x/s
  }
\end{listing}
This snippet scales elements in a vector \code{v1} relative to the sum
of \code{v2}'s elements. Without any extra work, the generic code motion
transform places the calculation of \code{s} (which is itself a loop) outside
the loop over \code{v1} because it
does not depend on the loop variable \code{x}.
\begin{listing}
  val s = sum(v2.length) { i => v2(i) }
  v1 map { x =>
    x/s
  }
\end{listing}

To ensure that operations can be safely moved around (and for other optimizations as well),
a compiler needs to reason about their possible side effects. Of particular interest is
the treatment of mutable data structures. Our current model, which works reasonably
well for the applications we have been studying so far (but might be overly
restrictive for others) is to make DSL authors annotate IR nodes with the
kind of effects they incur and prohibit sharing and aliasing between
mutable objects. Furthermore, read and write operations must unambiguously
identify the allocation site of the object being accessed.

By default, DSL operations are assumed pure (i.e.\ side-effect free).
DSL developers designate effectful operations using one of several \code{reflect} methods.
Console output, for example is implemented like this:
\begin{listing}
  def print(x: Exp[String]): Exp[Unit] = reflect(Print(x))
\end{listing}

The call to \code{reflect} adds the passed IR node to a list of effects for the current block. 
Effectful expressions will have dependency edges between them to ensure serialization. 
A compound expression such as a loop will internally use \code{reflect}'s counterpart,
called \code{reifyEffects}, to access the effectful expressions of its loop body. 
Effectful statements are tagged with an effect summary that further describes the effect.
The summary can be extracted via \code{summarizeEffects}, and 
there are some operations on summaries (\code{orElse}, \code{andThen}) to combine effects.
As an example consider the definition of conditionals, which computes the
compound effect from the effects of the two branches:
\begin{listing}
def __ifThenElse[T](cond: Exp[Boolean], thenp: => Rep[T], elsep: => Rep[T]) {
  val a = reifyEffects(thenp)
  val b = reifyEffects(elsep)
  val ae = summarizeEffects(a) // get summaries of the branches
  val be = summarizeEffects(b) 
  val summary = ae orElse be   // compute summary for whole expression
  reflectEffect(IfThenElse(cond, a, b), summary)  // reflect compound expression
                                                  // (effect might be none, i.e. pure)
}
\end{listing}

Up to here, we have encountered only binary effects: Either an operation has a global
effect (like \code{print}) or not. For reasoning about mutable data we clearly need something
more fine grained. To that end, we add further \code{reflect} methods:
\begin{listing}
  reflect           // a 'simple' effect: serialized with other simple effects
  reflectMutable    // an allocation of a mutable object. result guaranteed unique
  reflectWrite(v)   // a write to v: must refer to a mutable allocation (reflectMutable IR node)
  reflectRead(v)    // a read of allocation v (not used by programmer, inserted implicitly)
  reflectEffect(s)  // provide explicit summary s, specify may/must info for multiple reads/writes
\end{listing}

The framework will serialize reads and writes so to respect data and anti-dependency with respect to the referenced allocations.
To make this work we also need to keep track of sharing and aliasing. DSL authors can provide for their IR nodes 
a list of input expressions which the result of the IR node may alias, contain, extract from or copy from. 
These four pieces of information correspond to the possible pointer 
operations \code{x = y}, \code{*x = y}, \code{x = *y} and \code{*x = *y}. Using this knowledge, the framework
prohibits sharing between mutable objects and keeps track of immutable objects that point to mutable data. 
This is to make sure the right serialization dependencies and \code{reflectRead} calls are inserted for
operations that may indirectly reference mutable state.

This system seems to work well for programs that use a dominantly functional style 
but the no-sharing policy might be too restrictive for programs that make more extensive 
use of mutation. 
Effect systems and analyses are a large research topic on their own so we expect that further
research is needed and more DSLs and applications need to be studied.
Fortunately, Delite is not inherently tied to this particular effect system.

\subsection{DSL Programs are Program Generators}

LMS is a dynamic multi-stage programming approach: We have the full Scala language
at our disposal to compose fragments of DSL code. In fact, DSL programs are program
\emph{generators} that produce the DSL IR when run. 
DSL authors and application programmers can exploit this multi-level nature
to perform computations explicitly at staging time, so that the generated program does
not pay a runtime cost. 
Multi-stage programming thus shares some similarities with partial evaluation \cite{jones1993partial},
but instead of an automatic binding-time analysis, the programmer makes
binding times explicit in the program.
LMS uses \code{Rep} types for this purpose:
\begin{listing}
  val s: Int = ...            // a static value: computed at staging time
  val d: Rep[Int] = ...       // a dynamic value: computed when generated program is run
\end{listing}

Unlike with automatic partial evaluation, the programmer obtains a guarantee about 
which expressions will be evaluated at staging time.

While moving \emph{computations} from run time to staging time is an interesting
possibility, many computations actually depend on dynamic input and cannot be done
before the input is available (we will consider optimization of dynamic expressions below in Section~\ref{subsec:data}).
Nonetheless, explicit staging can be used to \emph{combine} dynamic computations more
efficiently.
Modern programming languages provide indispensable constructs for abstracting and
combining program functionality. Without higher-order features such as first-class
functions or object and module systems, software development at scale would not be possible. 
However, these abstraction mechanisms have a cost and make it much harder for the 
compiler to generate efficient code.

Using explicit staging, we can use abstraction in the generator stage to remove 
abstraction in the generated program. This holds both for control (e.g.\ functions, continuations)
and data abstractions (e.g.\ objects, boxing).

%\subsection{Control Abstractions}

\subsubsection{Leveraging Higher-Order Functions in the Generator}

Higher-order functions are extremely useful to structure programs but also
pose a significant obstacle for compilers, recent
advances on higher-order control-flow analysis notwithstanding \cite{DBLP:conf/esop/VardoulakisS10,DBLP:journals/corr/abs-1007-4268}.
While we would like to retain the structuring aspect for DSL programs,
we would like to avoid higher-order control flow in generated code. 
Fortunately, we can use higher-order functions in the generator stage to
compose first-order DSL programs.

Consider the following program that prints the number of elements greater than 7 in some vector:
\begin{listing}
  val xs: Rep[Vector[Int]] = ...
  println(xs.count(x => x > 7))
\end{listing}

The program makes essential use of a higher-order function \code{count} to
count the number of elements in a vector that fulfill a predicate given as
a function object. 
Ignoring
for the time being that we would likely want to use a \code{DeliteOp} for
parallelism, a good and natural way to implement \code{count} would be to
first define a higher-order function \code{foreach} to iterate over vectors:
\begin{listing}
  def infix_foreach[A](v: Rep[Vector[A]])(f: Rep[A] => Rep[Unit]) = {
    var i: Rep[Int] = 0
    while (i < v.length) {
      f(v(i))
      i += 1
    }
  }
\end{listing}

The actual counting can then be implemented in terms of the traversal:
\begin{listing}
  def infix_count[A](v: Rep[Vector[A]])(f: Rep[A] => Rep[Boolean]) = {
    var c: Rep[Int] = 0
    v foreach { x => if (f(x)) c += 1 }
    c
  }
\end{listing}

It is important to note that \code{infix_foreach} and \code{infix_count}
are static methods, i.e.\ calls will happen at staging time and result
in inserting the computed DSL code in the place of the call.
Likewise, while the argument vector \code{v} is a dynamic
value, the function argument \code{f} is again static. However, \code{f} operates
on dynamic values, as made explicit by its type \code{Rep[A] => Rep[Boolean]}.
By contrast, a dynamic function value would have type \code{Rep[A => B]}.

This means that the code generated for the example program will look roughly
like this, assuming that vectors are represented as arrays in the generated code:

\begin{listing}
  val v: Array[Int] = ...
  var c = 0
  var i = 0
  while (i < v.length) {
    val x = v(i)
    if (x > 7)
      c += 1
    i += 1
  }
  println(c)
\end{listing}

All traces of higher-order control flow have been removed and the program is
strictly first-order. Moreover, the programmer can be sure that the DSL program
is composed in the desired way.

This issue of higher-order functions is a real problem for regular Scala
programs executed on the JVM. The Scala collection library uses essentially the
same \code{foreach} and count \code{abstractions} as above and the JVM, which applies 
optimizations based on per-call-site profiling, will identify the call site \emph{within} \code{foreach} as
a hot spot. However, since the number of distinct functions called from foreach is
usually large, inlining or other optimizations cannot be applied and every iteration
step pays the overhead of a virtual method call \cite{cliffinlining}.

\subsubsection{Using Continuations in the Generator to Implement Backtracking}

Apart from pure performance improvements, we can use functionality of the generator stage
to enrich the functionality of DSLs without any work on the DSL-compiler side. As an example
we consider adding backtracking nondeterministic computation to a DSL using a simple variant of
McCarthy's \code{amb} operator \cite{McCarthy63abasis}. Here is a nondeterministic program that uses \code{amb}
to find pythagorean triples from three given vectors:
\begin{listing}
  val u,v,w: Rep[Vector[Int]] = ...
  nondet {
    val a = amb(u)
    val b = amb(v)
    val c = amb(w)
    require(a*a + b*b == c*c)
    println("found:")
    println(a,b,c)
  }
\end{listing}

We can use Scala's support for delimited continuations \cite{DBLP:conf/icfp/RompfMO09}
and the associated control operators \code{shift} and \code{reset} \cite{danvy1992rcs,Danvy90abstractingcontrol} to implement
the necessary primitives. 
The scope delimiter \code{nondet} is just the regular \code{reset}. The other operators are defined 
as follows:

\begin{listing}
  def amb[T](xs: Rep[Vector[T]]): Rep[T] @cps[Rep[Unit]] = shift { k =>
    xs foreach k
  }  
  def require(x: Rep[Boolean]): Rep[Unit] @cps[Rep[Unit]] = shift { k => 
    if (x) k() else ()
  }
\end{listing}

Since the implementation of \code{amb} just calls the previously defined method \code{foreach}, the
generated code will be first-order and consist of three nested \code{while} loops:
\begin{multicols}{2}
\begin{listing}
  val u,v,w: Rep[Vector[Int]] = ...
  var i = 0
  while (i < u.length) {
    val a = u(i)
    val a2 = a*a
    var j = 0
    while (j < v.length) {
      val b = v(j)
      val b2 = b*b
      val a2b2 = a2+b2
      var k = 0
      while (k < w.length) {
        val c = w(k)
        val c2 = c*c
        if (a2b2 == c2) {
          println("found:")
          println(a,b,c)
        }
        k += 1
      }
      j += 1
    }
    i += 1
  }
\end{listing}
\end{multicols}

Besides the advantage of not having to implement \code{amb} as part of the DSL compiler,
all common optimizations that apply to plain \code{while} loops are automatically applied to the
unfolded backtracking implementation. For example, note how loop invariant hoisting has
moved the computation of \code{a*a} and \code{b*b} out of the innermost loop.

The given implementation of \code{amb} is not the only possibility, though. For
situations where we know the number of choices (but not necessarily the actual values) for 
a particular invocation of \code{amb} at staging time, we can implement an 
alternative operator that takes a (static) list of
dynamic values and unfolds into specialized code paths for each option at compile
time:

\begin{listing}
  def bam[T](xs: List[Rep[T]]): Rep[T] @cps[Rep[Unit]] = shift { k =>
    xs foreach k
  }
\end{listing}

Here, \code{foreach} is not a DSL operation but a plain traversal of the static
argument list xs. The \code{bam} operator must be employed with some care because
it incurs the risk of code explosion. However, static specialization of 
nondeterministic code paths can be beneficial if it allows aborting many paths early
based on static criteria or merging computation between paths.

\begin{listing}
  val u: Rep[Vector[Int]] = ...
  nondet {
    val a = amb(u)
    val b = bam(List(x1), List(x2))
    val c = amb(v)
    require(a + c = f(b))  // assume f(b) is expensive to compute
    println("found:")
    println(a,b,c)
  }
\end{listing}

If this example was implemented as three nested loops, \code{f(x1)} and \code{f(x2)} would
need to be computed repeatedly in each iteration of second loop as they depend on the
loop-bound variable \code{b}. However, the use of \code{bam} will
remove the loop over \code{x1,x2} and expose the expensive computations as
redundant so that code motion can extract them from the loop:

\begin{listing}
  val fx1 = f(x1)
  val fx2 = f(x2)
  while (...) { // iterate over u
    while (...) { // iterate over v
      if (a + c == fx1) // found
    }
    while (...) { // iterate over v
      if (a + c == fx2) // found
    }
  }
\end{listing}

In principle, the two adjacent inner loops could be subjected to the loop fusion optimization discussed
in Section~\ref{subsec:fusion}. This would remove the duplicate traversal of \code{v}.
In this particular case fusion is currently not applied since it would change the order of the
side-effecting \code{println} statements.

%Second example: Translate use of control operators (e.g. McCarthy's amb) to first-order programs.
%Implement control operators using first-class continuations in the generator, then unfold 
%specialized code branches (may get code explosion though).

\subsection{Data Objects}
\label{subsec:data}

Besides control abstraction, the overhead of data abstraction is a major concern for performance
oriented programs. As a running example we consider adding a complex number datatype to our DSL.
The usual approach of languages executed on the JVM is to represent every non-primitive value as 
a heap-allocated reference object. The space overhead, reference indirection as well as the 
allocation and garbage collection cost are a burden for performance critical code.
Thus, we want to be sure that our complex numbers can be manipulated as efficiently
as two individual doubles. In the following, we explore different ways to achieve that.

%\note{mention Guy Steele's growing a language? Adding virtualization requirement: must be as efficient as built-in!}

%Example: Complex Numbers / Scalar Replacement. Let us assume we want to implement complex arithmetic 
%without overhead (Guy Steele's growing a language. Adding virtualization requirement: must be as efficient as built-in!)

%Goal: ensure we never created boxed Complex objects at program run time (only at staging time).

\subsubsection{Variant A: Static Data Structure}
\label{subsubsec:complexA}

The simplest approach is to implement complex numbers as a fully static data type, that
only exists at staging time. Only the actual \code{Double}s that constitute the
real and imaginary components of a complex number are dynamic values:

\begin{listing}
  case class Complex(re: Rep[Double], im: Rep[Double])
  def infix_+(a: Complex, b: Complex) = Complex(a.re + b.re, a.im + b.im)
  def infix_*(a: Complex, b: Complex) = Complex(a.re*b.re - a.im*b.im, a.re*b.im + a.im*b.re)
\end{listing}

Given two complex numbers \code{c1,c2}, an expression like
\begin{listing}
  c1 + 5 * c2  // assume implicit conversion from Int to Complex
\end{listing}
will generate code that is free of \code{Complex} objects and only contains arithmetic 
on \code{Double}s.

However the ways we can use \code{Complex} objects are rather limited. Since they
only exists at staging time we cannot, for example, express dependencies on dynamic
conditions:

\begin{listing}
  val test: Rep[Boolean] = ...
  val c3 = if (test) c1 else c2 // type error: c1/c2 not a Rep type
\end{listing}

It is worthwhile to point out that nonetheless, purely static data structures
have important use cases. To give an example, the fast fourier transform (FFT) \cite{cooley1965algorithm}
is branch-free for a fixed input size. The definition of complex numbers
given above can be used to implement a staged FFT that computes the well-known
butterfly shaped computation circuits from the textbook Cooley-Tukey recurrences 
\cite{DBLP:conf/emsoft/KiselyovST04,rompf10lms}.

To make complex numbers work across conditionals, 
we have have to split the control flow explicitly (another option would be
using mutable variables). 
There are multiple ways to achieve this splitting. 
We can either duplicate the test and create a single
result object:
  
\begin{listing}
  val test: Rep[Boolean] = ...
  val c3 = Complex(if(test) c1.re else c2.re, if(test) c1.im else c2.im)
\end{listing}

Alternatively we can use a single test and duplicate the rest of the program:

\begin{listing}
  val test: Rep[Boolean] = ...
  if (test) {
    val c3 = c1
    // rest of program
  } else {
    val c3 = c2
    // rest of program
  }
\end{listing}

While it is awkward to apply this transformation manually, we can use continuations (much like
for the \code{bam} operator) to generate two specialized computation paths:

\begin{listing}
  def split[A](c: Rep[Boolean]) = shift { k: (Boolean => A) =>
    if (c) k(true) else k(false) // "The Trick"
  }
  val test: Rep[Boolean] = ... 
  val c3 = if (split(test)) c1 else c2
\end{listing}

The generated code will be identical to the manually duplicated, specialized version above. 

\subsubsection{Variant B: Dynamic Data Structure with Partial Evaluation}

We observe that we can increase the amount of statically possible computation (in a sense,
applying binding-time improvements) for dynamic values with domain-specific rewritings:
\begin{listing}
  val s: Int = ...            // static  
  val d: Rep[Int] = ...       // dynamic

  val x1 = s + s + d          // left assoc: s + s evaluated statically, one dynamic addition
  val x2 = s + (d + s)        // naively: two dynamic additions, using pattern rewrite: only one
\end{listing}

In computing \code{x1}, there is only one dynamic addition because the left associativity of
the plus operator implies that the two static values will be added together at staging time.
Computing \code{x2} will incur two dynamic additions, because both additions have at least
one dynamic summand. However we can add rewriting rules that first replace \code{d+c}
(\code{c} denoting a dynamic value that is know to be a static constant, i.e.\ an IR
node of type \code{Const}) with \code{c+d} and then \code{c+(c+d)} with \code{(c+c)+d}.
The computation \code{c+c} can again be performed statically.

In a similar spirit, we can define a framework for data structures:
\begin{listing}
  trait StructExp extends BaseExp {
    case class Struct[T](tag: String, elems: Map[String,Rep[Any]]) extends Def[T]
    case class Field[T](struct: Rep[Any], index: String) extends Def[T]

    def field[T](struct: Rep[Any], index: String): Rep[T] = struct match {
      case Def(Struct(tag, elems)) => elems(index).asInstanceOf[Rep[T]]
      case _ => Field[T](struct, index)
    }
  }
\end{listing}
There are two IR node types, one for structure creation and one for field access.
The structure creation node contains a hash map that holds (static) field identifiers
and (dynamic) field values.
The interface for field accesses is method \code{field}, which pattern matches
on its argument and, if that is a \code{Struct} creation, looks up the desired value
from the embedded hash map.

An implementation of complex numbers in terms of \code{Struct} could look like this:
\begin{listing}
  trait ComplexOps extends ComplexBase with ArithOps {
    def infix_+(x: Rep[Complex], y: Rep[Complex]): Rep[Complex] = Complex(x.re + y.re, x.im + y.im)
    def infix_*(x: Rep[Complex], y: Rep[Complex]): Rep[Complex] = Complex(a.re*b.re - ...)
  }
  trait ComplexBase extends Base {
    class Complex
    def Complex(re: Rep[Double], im: Rep[Double]): Rep[Complex]
    def infix_re(c: Rep[Complex]): Rep[Double]
    def infix_im(c: Rep[Complex]): Rep[Double]
  }
  trait ComplexStructExp extends ComplexBase with StructExp {
    def Complex(re: Rep[Double],im: Rep[Double])=Struct[Complex]("Complex", Map("re"->re, "im"->im))
    def infix_re(c: Rep[Complex]): Rep[Double] = field[Double](c, "re")
    def infix_im(c: Rep[Complex]): Rep[Double] = field[Double](c, "im")
  }
\end{listing}

Note how complex arithmetic is defined completely within the interface trait \code{ComplexOps},
which inherits double arithmetic from \code{ArithOps}. Access to the components via
\code{re} and \code{im} is implemented using \code{Struct}.

In contrast to the completely static implementation of complex numbers presented in 
Section~\ref{subsubsec:complexA} above, complex numbers are a fully dynamic
DSL type now. The previous restrictions are gone and we can write the following
code without compiler error:

\begin{listing}
  val c3 = if (test) c1 else c2
  println(c3.re)
\end{listing}

However there is still one ingredient missing. Taking only the implementation of \code{Struct}
seen so far, the conditional that computes \code{c3} would need to create some kind of
a \code{StructDyn} object at runtime, from which the invocation of \code{re} would
then need to retrieve the stored data. After all, the implementation of \code{field} can
only lookup the field statically if the argument is a \code{Struct}, not an
\code{IfThenElse} node. What is missing is thus a rule that makes the result 
of a conditional a \code{Struct} if the branches return \code{Struct}:

\begin{listing}
  override def ifThenElse[T](cond: Rep[Boolean], a: Rep[T], b: Rep[T]) = (a,b) match {
    case (Def(Struct(tagA,elemsA)), Def(Struct(tagB, elemsB))) => 
      assert(tagA == tagB)
      assert(elemsA.keySet == elemsB.keySet)
      Struct(tagA, for (k <- elemsA.keySet) yield (k -> ifThenElse(cond, elemsA(k), elemsB(k))))
    case _ => super.ifThenElse(cond,a,b)
  }
\end{listing}

Similar rules are added for many of the other core IR node types.

There is another particularly interesting use case: Let us assume we want to create a vector of
complex numbers. Just as with the if then else example above, we can override the vector
constructors such that a \code{Vector[Complex]} is represented as a struct that contains
two separate arrays, one for the real and one for the imaginary components.
In fact, we have expressed our conceptual array of structs as a struct of arrays.
This data layout is beneficial in many cases. Consider for example calculating complex
conjugates (i.e.\ swapping the sign of the imaginary compoents) over a vector of complex numbers.
All the real parts remain unchanged so the array holding them need not be touched at all.
Only the imaginary parts have to be transformed, cutting the total required memory bandwidth
in half. Moreover, uniform array operations like this are a much better fit for SIMD execution.

We conclude this section by taking note that we can actually guarantee that no dynamic \code{Complex} or 
\code{Struct} object is ever created just by not implementing code generation logic for \code{Struct}
and \code{Field} IR nodes and signaling an error instead. This is a good example of a performance-oriented
DSL compiler rejecting a program as ill-formed because it cannot be executed in the desired,
efficient way.

\subsection{Extending the Framework}
A framework for building DSLs must be easily extensible in order for the DSL developer to exploit domain
knowledge starting from a general-purpose IR design.  Consider a simple DSL for linear algebra with a 
Vector type.  Now we want to add norm and dist functions to the DSL. The first possible implementation
is to simply implement the functions as library methods.

\begin{listing}
  def norm[T:Numeric](v: Rep[Vector[T]]) = {
    sqrt(v.map(j => j*j).sum)
  }
  def dist[T:Numeric](v1: Rep[Vector[T]], v2: Rep[Vector[T]]) = {
    norm(v1 - v2)
  }
\end{listing}

Whenever the dist method is called the implementation will be added to the application IR in terms of vector subtraction,
vector map, vector sum, etc. (assuming each of these methods is built-in to the language rather than also being provided
as a library method).  This version is very straightforward to write but the knowledge that the application wishes to 
find the distance between two vectors is lost.

By defining norm explicitly in the IR implementation trait (where Rep[T] = Exp[T]) we gain ability to perform pattern matching
on the IR nodes that compose the arguments.

\begin{listing}
  override def norm[T:Numeric](v: Exp[Vector[T]]) = v match {
    case Def(ScalarTimesVector(s,u)) => s * norm(u)
    case Def(ZeroVector(n)) => 0
    case _ => super.norm(v)
  }
\end{listing}

In this example there are now three possible implementations of \code{norm}.  The first case factors scalar-vector multiplications out 
of \code{norm} operations, the second short circuits the norm of a ZeroVector to be simply the constant 0, and the third falls back 
on the default implementation defined above.  With this method we can have a different implementation of \code{norm} for each 
\emph{occurrence} in the application.

An even more powerful alternative is to implement \code{norm} and \code{dist} as custom IR nodes.  This enables the DSL to include these nodes
when optimizing the application via pattern matching and IR rewrites as illustrated above.  For example, we can add a rewrite
rule for calculating the norm of a unit vector: if  $v_1 = \frac{v}{\|v\|}$ then $\left\|v_1\right\|=1$.
In order to implement this optimization we need to add cases both for the new \code{norm} operation as well as to the
existing scalar-times-vector operation to detect the first half of the pattern.

\begin{listing}
  case class VectorNorm[T](v: Exp[Vector[T]]) extends Def[T]
  case class UnitVector[T](v: Exp[Vector[T]]) extends Def[Vector[T]]
  
  override def scalar_times_vector[T:Numeric](s: Exp[T], v: Exp[Vector[T]]) = (s,v) match {
    case (Def(Divide(Const(1), Def(VectorNorm(v1)))), v2) if v1 == v2 => UnitVector(v)
    case _ => super.scalar_times_vector(s,v)
  }
  override def norm[T:Numeric](v: Exp[Vector[T]]) = v match {
    case Def(UnitVector(v1)) => 1
    case _ => super.norm(v)
  }
\end{listing}

\comment{
\begin{listing}
  override def dist[T:Numeric](v1: Exp[Vector[T]], v2: Exp[Vector[T]]) = (v1, v2) match {
    case same if (v1 == v2) => 0
    case (v, Def(ScalarTimesVector(n@Def(Norm(v)), v))) => abs(n-1)*n
    case (Def(ScalarTimesVector(n@Def(Norm(v)), v)), v) => abs(n-1)*n
    case (v, Def(ScalarTimesVector(s,v)) => abs(s-1)*norm(v)
    case (Def(ScalarTimesVector(s,v)), v) => abs(s-1)*norm(v)
    case _ => super.dist(v1, v2)
  }
\end{listing}
}

In this example the scalar-times-vector optimization requires vector-norm to exist as an IR node to detect\footnote{The \code{==}
operator tests structural equality of IR nodes.
The test is cheap because we only need to look at symbols, one level deep. 
Value numbering/CSE ensures that intensionally equal IR nodes get assigned the same symbol.}  and short-circuit 
the operation to simply create and mark unit vectors.  The vector-norm optimization then detects unit vectors and short circuits the norm operation
to simply add the constant 1 to the IR.  In every other case it falls back on the default implementation, which is to create a new \code{VectorNorm} IR node.

When these domain-specific IR nodes (ScalarTimesVector, Norm, etc.) are created by extending DeliteOp nodes parallel code generation, generic
optimizations, and parallel optimizations are performed automatically by the framework.  In the case of norm and dist, an extremely useful 
performance optimization that the framework provides is fusing the individual operations so that the distance is computed with a single pass 
over the two vectors rather than the three passes that would occur with a straightforward generation of dist as written.  We will now look
at the loop fusion support provided by the framework in more detail.

\subsection{Fusion}
\label{subsec:fusion}
Building complex bulk operations out of simple ones often leads to inefficient generated code.  For example consider the simple vector code

\begin{listing}
  val a: Rep[Double] = ...
  val x: Rep[Vector[Double]] = ...
  val y: Rep[Vector[Double]] = ...
  
  a*x+y
\end{listing} 

Assuming we have provided the straightforward loop-based implementations of scalar-times-vector and vector-plus-vector, the resulting code for
this program will perform two loops and allocate a temporary vector to store \code{a*x}.  A more efficient implementation will only use
a single loop (and no temporary vector allocations) to compute \code{a*x(i)+y(i)}.

In addition to operations that are directly dependent as illustrated above, side-by-side operations also appear frequently.
As an example, consider a DSL that provides mean and variance methods.

\begin{listing}
  def mean(x: Rep[Vector[Double]]) = 
      sum(x.length) { i => x(i) } / x.length
  def variance(x: Rep[Vector[Double]]) =
      sum(x.length) { i => square(x(i)) } / x.length - square(mean(x))
  
  val data = ...
  
  val m = mean(data)
  val v = variance(data)
\end{listing}

%\note{TODO: how are dependencies specified}

The DSL developer wishes to provide these two functions separately, but many applications will compute both the mean and variance of a 
dataset together.  In this case we again want to perform all the work with a single pass over \code{data}.  In both of the above example situations,
fusing the operations into a single loop greatly improves cache behavior and reduces the total number of loads and stores required. 
It also creates coarser-grained functions out of fine-grained ones, which will likely improve parallel scalability.  

Our framework handles all situations like these two examples uniformly and for all DSLs.  Any non-effectful IR node that extends DeliteOpLoop 
is eligible for fusing with other DeliteOpLoops.  In order to handle all the interesting loop fusion cases, the fusing algorithm uses
a simple and general criterion: It fuses all pairs of loops where either both loops have the exact same size or one loop iterates over
a data structure the other loop creates, as long as fusing will not create any cyclic dependencies.
When it finds two eligible loops the algorithm creates a new loop with a body composed of both of the original bodies.  
Merging loop bodies includes array contraction, i.e.\ the fusing transform modifies dependencies so that all results produced 
within a loop iteration are consumed directly rather than by reading an output data structure.
Whenever this renders an output data structure unnecessary (it does not escape the fused loop) it is removed automatically by the dead code elimination 
system.  All DeliteOpLoops are parallel loops, which allows the fused loops to be parallelized in the same manner as the original loops.     

The general heuristic is to apply fusion greedily wherever possible. For dominantly imperative code more 
refined heuristics might be needed \cite{DBLP:conf/sc/BelterJKS09}. 
However, loop abstractions in Delite are dominantly functional and
many loops create new data structures. Removing intermediate data buffers,
which are potentially large and many of which are used only once is clearly a win, 
so fusing seems to be beneficial in almost all cases.

Our fusion mechanism is similar but not identical to deforestation \cite{DBLP:journals/tcs/Wadler90} and related 
approaches \cite{DBLP:conf/icfp/CouttsLS07}. 
Many of these approaches only consider expressions that are directly dependendent, whereas we
are able to handle both dependent and side-by-side expressions with one general mechanism.  This is critical for situations such as the
mean and variance example, where the only other efficient alternative would be to explicitly create a composite function that returns
both results simultaneously.  This solution additionally requires the application writer to always remember to use the composite version 
when appropriate.  It is generally difficult to predict all likely operation compositions as well as onerous to provide efficient, specialized 
implementations of them.  Therefore fusion is key for efficient compositionality in both applications and DSL libraries.

%% file: optiml.tex
\section{Case Study: OptiML}

OptiML is an embedded DSL for machine learning (ML) that we have developed on
top of LMS and Delite.  It provides a MATLAB-like programming model with
ML-specific abstractions. OptiML is a prototypical example of how the techniques
described in this paper can be used to construct productive, high performance
DSLs targeted at heterogeneous parallel machines. 

\label{sec:optiml}
%\subsection{Streaming Matrix}
\subsection{Downsampling in Bioinformatics}

In this example, we will demonstrate how the optimization and code generation
techniques discussed in previous sections come together to produce efficient
code in real applications. SPADE is a bioinformatics application
that builds tree representations of large, high-dimensional flow cytometry datasets.
Consider the following small but compute-intensive snippet from SPADE (C++):

\begin{listing}
  std::fill(densities, densities+obs, 0);
  #pragma omp parallel for shared(densities)  
  for (size_t i=0; i<obs; i++) {
    if (densities[i] > 0)
      continue;
    std::vector<size_t> apprxs;  // Keep track on observations we can approximate
    Data_t *point = &data[i*dim];
    Count_t c = 0;

    for (size_t j=0; j<obs; j++) {
      Dist_t d = distance(point, &data[j*dim], dim);
      if (d < apprx_width) {
        apprxs.push_back(j);
        c++;
      } else if (d < kernel_width) c++;
    }
    // Potential race condition on other density entries, use atomic
    // update to be safe
    for (size_t j=0; j<apprxs.size(); j++)
      __sync_bool_compare_and_swap(densities+apprxs[j],0,c); //densities[apprxs[j]] = c;
    densities[i] = c;
}
\end{listing}

This snippet represents a downsampling step that computes a set of values,
densities, that represents the number of samples within a bounded distance
(kernel\_width) from the current sample. Furthermore, any distances within
apprx\_width of the current sample are considered to be equivalent, and the
density for the approximate group is updated as a whole. Finally, the loop is
run in parallel using OpenMP. This snippet represents hand-optimized, high
performance, low-level code. It took a systems and C++ expert to port the
original MATLAB code (written by a bioinformatics researcher) to this
particular implementation. In contrast, consider the equivalent snippet of
code, but written in OptiML:

\begin{listing}
  val distances = Stream[Double](data.numRows, data.numRows){ (i,j) => dist(data(i),data(j)) }
  val densities = Vector[Int](data.numRows, true)

  for (row <- distances.rows) {
    if(densities(row.index) == 0) {
      val neighbors = row find { _ < apprxWidth }
      densities(neighbors) = row count { _ < kernelWidth }
    }
  }
  densities
\end{listing}

This snippet is expressive and easy to write. It is not obviously high
performance. However, because we have abstracted away implementation detail,
and built-in high-level semantic knowledge into the OptiML compiler, we can
generate code that is essentially the same as the hand-tuned C++ snippet. Let's
consider the OptiML code step by step.

Line 1 instantiates a Stream, which is an OptiML data structure that is
buffered; it holds only a chunk of the backing data in memory at a time, and
evaluates operations one chunk at a time. Stream only supports iterator-style
access and bulk operations. These semantics are necessary to be able to express
the original problem in a more natural way without adding overwhelming
performance overhead. The foreach implementation for stream.rows is:

\begin{listing}
  def stream_foreachrow[A:Manifest](x: Exp[Stream[A]], block: Exp[StreamRow[A]] => Exp[Unit]) = {
    var i = 0
    while (i < numChunks) {
      val rowsToProcess = stream_rowsin(x, i)
      val in = (0::rowsToProcess)
      val v = fresh[Int]

      // fuse parallel initialization and foreach function
      reflectEffect(StreamInitAndForeachRow(in, v, x, i, block))   // parallel
      i += 1
    }
  }
\end{listing}

This method constructs the IR nodes for iterating over all of the chunks in the
Stream, initalizing each row, and evaluating the user-supplied foreach
anonymous function. We first obtain the number of rows in the current chunk by
calling a method on the Stream instance (\code{stream_rowsin}). We then call
the StreamInitAndForeachRow op, which is a DeliteOpForeach, over all of the
rows in the chunk.  OptiML unfolds the foreach function and the stream
initialization function while building the IR, inside StreamInitAndForeachRow.
The stream initialization function (\code{(i,j) => dist(data(i),data(j)})
constructs a StreamRow, which is the input to the foreach function. The
representation of the foreach function consists of an IfThenElse operation,
where the then branch contains the VectorFind, VectorCount, and
VectorBulkUpdate operations from lines 6-7 of the OptiML SPADE snippet.
VectorFind and VectorCount both extend DeliteOpLoop. Since they are both
DeliteOpLoops over the same range with no cyclic dependencies, they are fused
into a single DeliteOpLoop. This eliminates an entire pass (and the
corresponding additional memory accesses) over the row, which is a non-trivial
235,000 elements in one typical dataset.

Fusion helps to transform the generated code into the iterative structure of
the C++ code. One important difference remains: we only want to compute the
distance if it hasn't already been computed for a neighbor. In the streaming
version, this corresponds to only evaluating a row of the Stream if the
user-supplied if-condition is true. In other words, we need to optimize the
initialization function \emph{together with} the anonymous function supplied to
the foreach. LMS does this naturally since the foreach implementation and the
user code written in the DSL are all uniformly represented with the same IR.
When the foreach block is scheduled, the stream initialization function is
pushed inside the user conditional because the StreamRow result is not required
anywhere else. Furthermore, once the initialization function is pushed inside
the conditional, it is then fused with the existing DeliteOpLoop, eliminating
another pass. We can go even further and remove all dependencies on the
StreamRow instance by bypassing field accesses on the row, using the pattern
matching mechanism described earlier:

\begin{listing}
  trait StreamOpsExpOpt extends StreamOpsExp {
    this: OptiMLExp with StreamImplOps =>
  
    override def stream_numrows[A:Manifest](x: Exp[Stream[A]]) = x match {
      case Def(Reflect(StreamObjectNew(numRows, numCols, chunkSize, func, isPure),_,_)) => numRows
      case _ => super.stream_numrows(x)
    }
    // similar overrides for other stream fields
  }
  trait VectorOpsExpOpt extends VectorOpsExp {
    this: OptiMLExp with VectorImplOps =>
    // accessing an element of a StreamRow directly accesses the underlying Stream
    override def vector_apply[A:Manifest](x: Exp[Vector[A]], n: Exp[Int]) = x match {
      case Def(StreamChunkRow(x, i, offset)) => stream_chunk_elem(x,i,n)
      case _ => super.vector_apply(x,n)
    }
  }
\end{listing}

Now as the row is computed, the results of VectorFind and VectorCount are also
computed in a pipelined fashion. All accesses to the StreamRow are
short-circuited to their underlying data structure (the Stream), and no
StreamRow object is ever allocated in the generated code. The following listing
shows the final code generated by OptiML for the 'then' branch (comments and
indentation added for clarity):

%val x164 = x158 * x64
%val x201 = {
%val x197 = new generated.scala.IndexVectorSeqImpl(0)
%x197
%}
%var x208: Int = 0
\begin{multicols}{2}
\begin{listing}
  // ... initialization code omitted ...
  // -- FOR EACH ELEMENT IN ROW --
  while (x155 < x61) {  
    val x168 = x155 * x64
    var x185: Double = 0
    var x180 = 0
  
    // -- INIT STREAM VALUE (dist(i,j))  --
    while (x180 < x64) {  
      val x248 = x164 + x180
      val x249 = x55(x248)
      val x251 = x168 + x180
      val x252 = x55(x251)
      val x254 = x249 - x252
      val x255 = java.lang.Math.abs(x254)
      val x184 = x185 + x255
      x185 = x184
      x180 += 1
    } 
    val x186 = x185
    val x245 = x186 < 6.689027961000001
    val x246 = x186 < 22.296759870000002
  
    // -- VECTOR FIND --
    if (x245) x201.insert(x201.length, x155)
  
    // -- VECTOR COUNT --
    if (x246) {
      val x207 = x208 + 1
      x208 = x207
    }
    x155 += 1
  } 

  // -- VECTOR BULK UPDATE --
  var forIdx = 0
  while (forIdx < x201.size) { 
    val x210 = x201(forIdx)
    val x211 = x133(x210) = x208
    x211
    forIdx += 1
  } 
\end{listing}
\end{multicols}

This code, though somewhat obscured by the compiler generated names, closely
resembles the hand-written C++ snippet shown earlier. It was generated from a
simple, 9 line description of the algorithm written in OptiML, making heavy use
of the building blocks we described in previous sections to produce the final
result.

%Discuss in detail how things work together to make it efficient.
%
%\begin{listing}
%case class Stream[A](val numRows: Int, val numCols: Int)(func: (Int,Int) => A) {
%  object rows {
%    
%    def foreach(f: StreamRow[A] => Unit): Unit = {
%      for (i <- 0 until numRows) {
%        val rowVec = new StreamRowImpl[A](i)
%        for (j <- 0 until numCols)
%          rowVec(j) = func(i,j)
%        f(rowVec)
%      }
%    }
%    
%  }
%  
%  
%  val streamA = Stream(200, 300) { (i,j) => dist(data(i), data(j)) }
%  val streamB = Stream("input.dat")
%  
%  val stream = if (bla) streamA else streamB
%\end{listing}

\subsection{Performance Measurements}

%\begin{figure*} \centering \includegraphics[clip=true,height=2in,trim=0in 0in
%0in 0in]{figures/spade.pdf} \caption{Normalized execution time of SPADE in C++
%and OptiML with and without fusing optimizations. Speedup numbers are reported
%on top of each bar.} \label{fig:spade} \end{figure*}
%
%\begin{figure*} \centering \includegraphics[clip=true,height=2in,trim=0in 0in
%0in 0in]{figures/tm.pdf} \caption{Normalized execution time of template
%matching in C++ and OptiML. Speedup numbers are reported on top of each bar.}
%\label{fig:tm} \end{figure*}

%\begin{figure*} \centering \includegraphics[clip=true,height=2in,trim=0in 0in
%0in 0in]{figures/gpu.pdf} \caption{Normalized execution time of 8-core CPU
%and GPU for a selection of applications in OptiML. Speedup numbers are reported
%on top of each bar.} \label{fig:gpu} \end{figure*}

%\begin{figure*}[tbh!]
%\centering
%\subfigure{
%  \includegraphics[width=0.33\linewidth]{figures/spade.pdf}
%      \includegraphics[width=0.33\linewidth]{figures/tm.pdf}
%          \includegraphics[width=0.33\linewidth]{figures/gpu.pdf}
%} 
%\end{figure*}

\begin{figure}[tbh!]
\vspace{-6mm}
\subfigure[Normalized execution time of SPADE in C++
and OptiML with and without fusing optimizations, for 1 to 8 CPU cores. 
Speedup relative to 1-core OptiML with fusing on top of each bar.]{ 
  \includegraphics[width=0.31\linewidth]{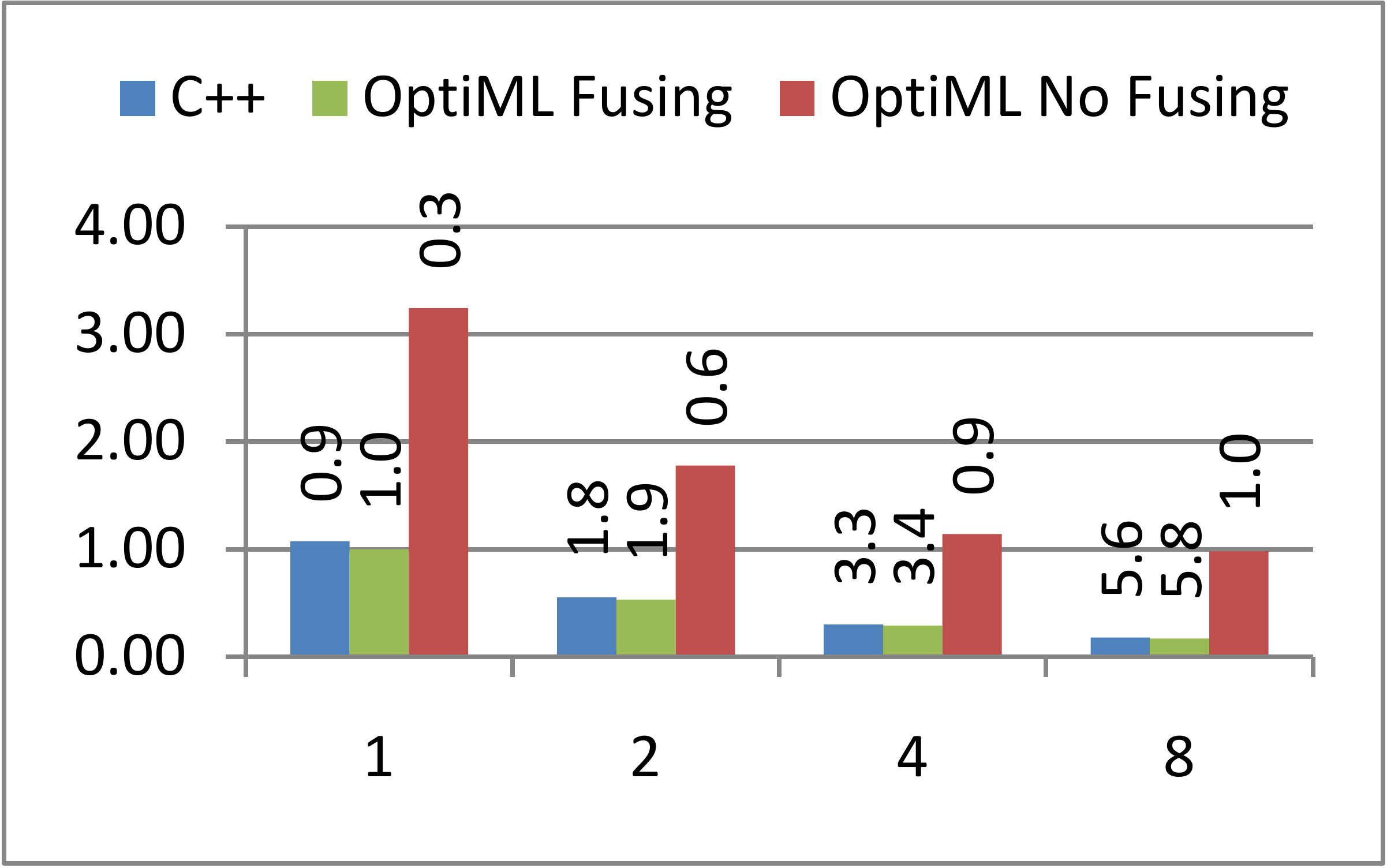}
  \label{fig:spade}
}\,\,
\subfigure[Normalized execution time of template matching in C++ and OptiML, for 1 to 8 CPU cores.
Speedup numbers relative to 1-core OptiML shown on top of each bar.]{ 
  \includegraphics[width=0.31\linewidth]{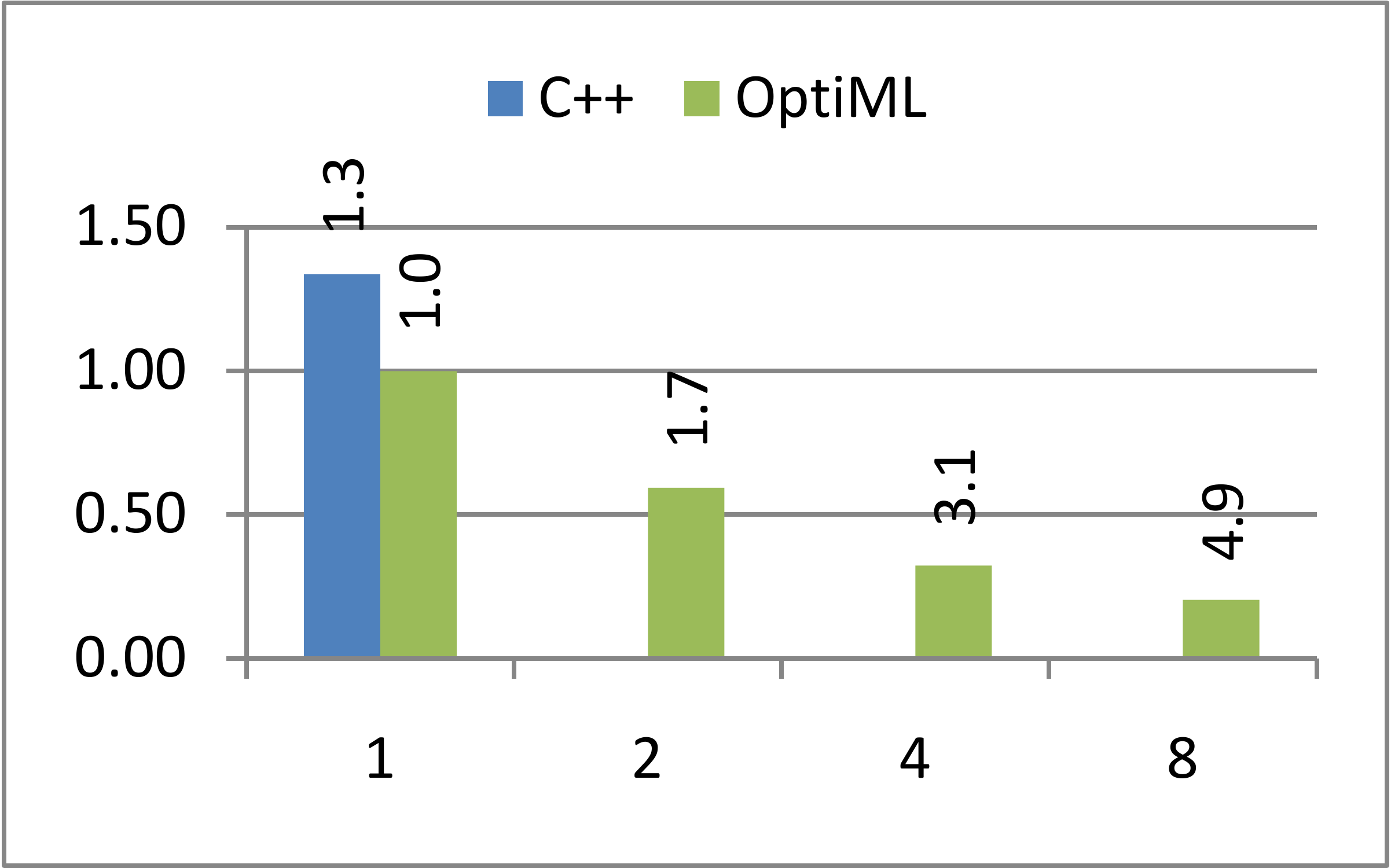}
  \label{fig:tm}
}\,\,
\subfigure[Normalized execution time of 8-core CPU and 1 GPU for a selection of applications in OptiML.
Speedup numbers relative to 1-core CPU shown on top of each bar.]{
  \includegraphics[width=0.31\linewidth]{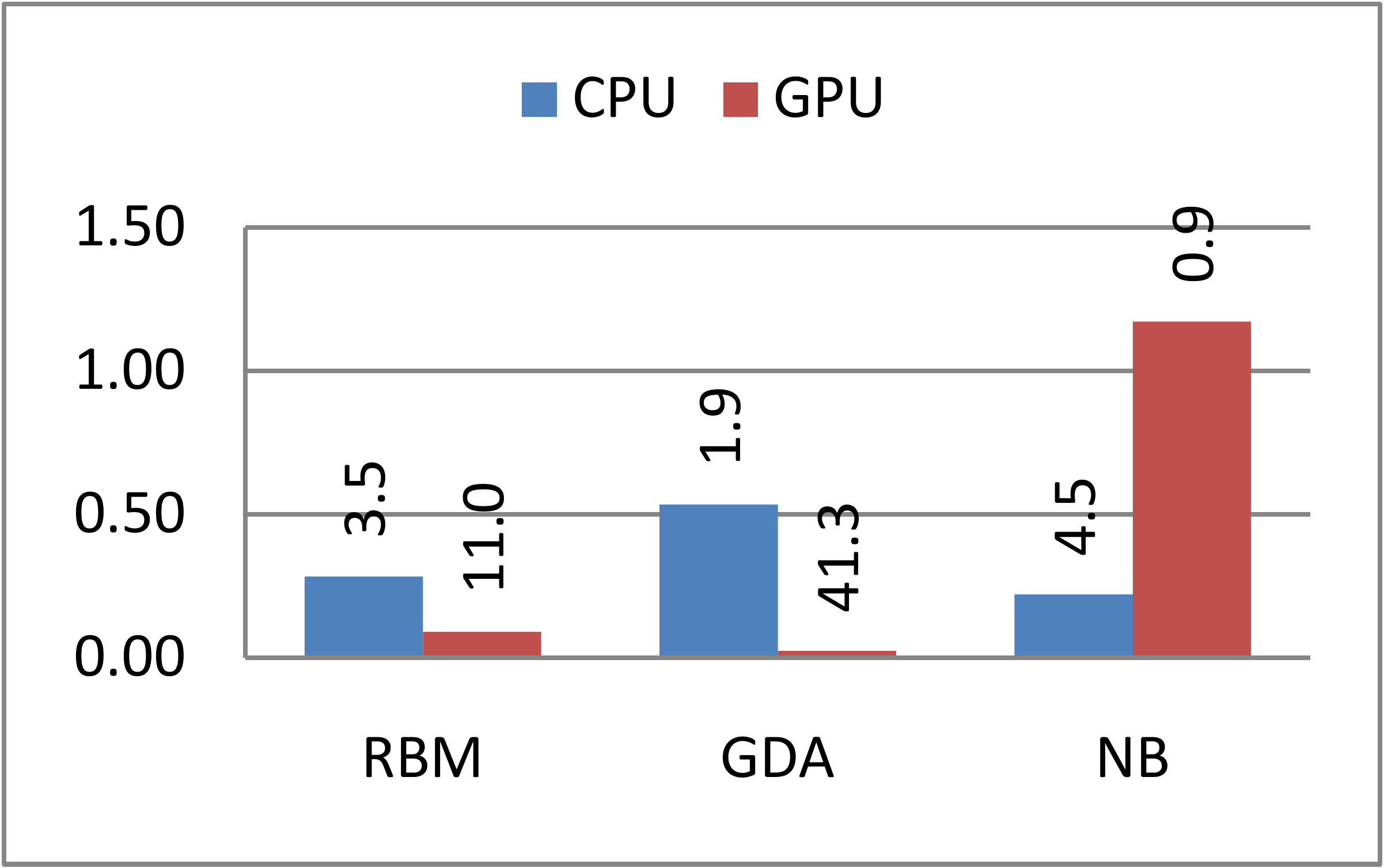}
  \label{fig:gpu}
}
\vspace{-1mm}
\caption{Performance results for the OptiML DSL running on Delite}
\end{figure} 

\vspace{-2mm}
Figures ~\ref{fig:spade} and ~\ref{fig:tm} show how OptiML performs relative to
hand-optimized C++ versions for two real-world applications: SPADE (discussed
in the previous section) and a template matching application \cite{bigg}
used for object recognition in robotics. For each experiment, we ran
each application 10 times and report the mean execution time of the final 5
executions (not counting initialization).  The experiments were run on a Dell
Precision T7500n with two quad-core Intel Xeon X5550 2.67 GHz processors, 24GB
of RAM, and an NVidia Tesla C2050. Our CPU results use generated Scala code,
compiled and executed with Oracle's Java SE Runtime Environment 1.7.0-b133 and
the HotSpot 64-bit server VM with default options.

The results show that OptiML generates code that performs comparably to, and
can even outperform, a hand-tuned C++ implementation. For SPADE, prior to
applying the optimizations discussed in the previous section, the code generated by
OptiML is three times slower than the C++ version. This overhead comes from
the extra memory allocations and loops over the data that are required in the
naive implementation. The naive version also does more computation than the C++,
because it always initializes a StreamRow even when it is not needed. 
By applying fusion and code motion as we described, the OptiML code becomes a tight loop
with no intermediate allocation. The small improvement over the C++ version
might be due to the JVM JIT producing slightly better native code and from the
removal of an atomic update, which is unnecessary on 64-bit platforms.

The OptiML version of the template mapping (TM) application is much shorter and
performs better than the C++ version, mostly due to removing a substantial
amount of low-level bit-manipulating optimizations from the application code
that did not perform as well on our platform. The C++ code also reused some
data structures in ways that made parallelizing non-trivial, while the OptiML
code was implicitly parallel from extending Delite Ops. Because the C++ TM is
sequential, we report only the single-threaded execution time; the OptiML
version scales efficiently to multiple processors. This application clearly
demonstrates that low-level implementation details in application code is a
liability to performance portability. 

Finally, Figure ~\ref{fig:gpu} shows our GPU performance for three
selected machine learning applications: Riemann Boltzmann Machine (RBM), Naive
Bayes (NB), and Gaussian Discriminant Analysis (GDA). The results shown are
relative to single-threaded OptiML performance. These results show that a
heterogeneous programming model is required to achieve best performance on
these applications; neither the CPU nor the GPU is best for all cases. The
OptiML code can run on either the CPU or GPU without any changes to the source
code. In previous work, we compared OptiML performance to MATLAB across a wider
range of applications, and found that we outperform MATLAB's CPU performance
and GPU performance in most cases \cite{icml11optiml}.

%% file: related.tex
\section{Related Work}
\label{sec:related}

Lightweight Modular Staging and Delite build upon previously published work in multiple areas, 
including DSLs, multi-stage compilation, and parallel programming.

\textbf{DSLs and multi-stage compilation:}
DSLs fall into two broad categories, namely external DSLs which are completely independent 
languages, and internal DSLs, which borrow functionality from a host language.  We use Hudak's
model of embedded DSLs \cite{hudak96building}.  Previous work has shown some of the benefits of 
domain-specific optimizations.  Guyver et al. present significant performance improvements by 
annotating library methods with domain-specific knowledge \cite{guyver99annotation}, and CodeBoost 
\cite{bagge03design} uses user-defined rules to transform programs using domain knowledge.

Multi-stage programming languages include MetaML \cite{DBLP:journals/tcs/TahaS00} and MetaOCaml 
\cite{DBLP:conf/gpce/CalcagnoTHL03}.  Several other static metaprogramming methods exist, 
including C++ templates \cite{vandevoorde2003} and Template Haskell \cite{sheardTemplateHaskell}.  
Expression Templates \cite{veldhuizen1996expression} can produce customized generation, and are 
used by Blitz++ \cite{DBLP:conf/iscope/Veldhuizen98}.  Veldhuizen introduced active libraries 
\cite{phd/Veldhuizen04}, which are libraries that participate in compilation.  Kennedy introduced 
telescoping languages \cite{kennedy05telescoping}, efficient DSLs created from annotated component 
libraries. TaskGraph \cite{beckmann04} is a meta-programming library that sports run-time code generation in C++.
Lightweight Modular Staging is built on the idea of embedding typed languages by 
Carette et al.\ \cite{DBLP:conf/aplas/CaretteKS07} and Hofer et al.\ \cite{hoffer08polymorphic}.  
Many existing program generators such as FFTW \cite{DBLP:conf/pldi/Frigo99}, ATLAS \cite{DBLP:journals/pc/WhaleyPD01}
and SPIRAL \cite{DBLP:journals/ijhpca/PuschelMSXJPVJ04} took enormous efforts to build.  LMS and 
Delite aim to make generative facilities more easily accessible.

\textbf{Heterogeneous parallel programming:}
Systems such as %EXOCHI \cite{wang07exochi} and 
OpenCL \cite{opencl} provide abstractions that 
allow the programmer to explicitly manage and target any available accelerator, eliminating the 
need to use vendor APIs for each device. Data-parallel programming models hide the complexity of 
the underlying hardware through an abstract data-parallel API.  Recent work in this area includes 
Copperhead \cite{catanzaro11copperhead}, which automatically generates Cuda code from a data-
parallel subset of Python, and FlumeJava \cite{chambers10flumejava}, which is a Java library that 
targets Google's MapReduce \cite{DBLP:conf/osdi/DeanG04} and optimizes the pipeline of MapReduce 
operations based on the data-flow graph. Intel's Array Building Blocks \cite{intelArBB} executes 
data-parallel patterns across processor cores and targets multiple architectures (e.g., different 
vector units) from a single application source. 
Concurrent Collections (CnC) \cite{DBLP:journals/sp/BudimlicBCKLNPPSST10} shares some similarities 
with the Delite task graph. Computation \emph{steps} in CnC are separate and opaque to the 
scheduling, whereas Delite produces optimized kernels that are well-integrated with the schedule. 
Recent work used embedded DSLs combined with a common parallel runtime to enable implicit task and 
data parallelism via deferred execution \cite{ppopp11delite}.  The DSLs, however, were unable to 
perform analyses and transformations of the applications.  

Several parallel programming languages exist, include Chapel \cite{chapel}, Fortress 
\cite{DBLP:conf/IEEEpact/Steele05}, and X10 \cite{x10}.  These languages require explicit control 
over locations and concurrency.  In contrast, the Delite runtime manages locations and concurrency 
transparently.  Implicit parallelism in languages is often based on data-parallel operations on 
parallel collections. Languages with this feature include Chapel, Data-Parallel Haskell 
\cite{DBLP:conf/fsttcs/JonesLKC08}, Fortress, High Performance Fortran \cite{hpf}, NESL \cite{nesl}, 
and X10.  DSLs which utilize the Delite framework are able exploit implicit data parallelism as well 
as implicit task parallelism.

\comment{
\textbf{on GPU DSLs:}

Vertigo\cite{elliott04vertigo}, Obsidian, Nikola \cite{mainland10nikola} and Accelerate \cite{chakravarty11accelerate} are embedded languages for GPGPU programming in Haskell.

Geoffrey Mainland and Greg Morrisett. Nikola: Embedding compiled GPU functions in Haskell. In Haskell ’10: Proceedings of the 2010 ACM SIGPLAN Symposium on Haskell. ACM, September 2010. \cite{mainland10nikola}

Accelerate (chakravarty), http://www.cse.unsw.edu.au/~chak/project/accelerate/, Haskell'09, DAMP'11 \cite{chakravarty11accelerate}

---
ScalaCL: transform Scala subset via compiler plugin. http://code.google.com/p/scalacl/ \cite{scalacl}

Thomas Jansen. GPU++: An Embedded GPU Development System for General-Purpose Computations. PhD thesis, Technische Universität München, 2008.

Andreas Klöckner, Nicolas Pinto, Yunsup Lee, Bryan C. Catanzaro, Paul Ivanov, and Ahmed Fasih. PyCUDA: GPU run-time code gener- ation for high-performance computing. CoRR, 2009. \cite{kloeckner09pycuda}

---

anything about IBM's Lime? Auerbach et al. OOPSLA'10. \cite{Auerbach10lime}

Guaranteed Optimization: Proving Nullspace Properties of Compilers
Todd L. Veldhuizen and Andrew Lumsdaine, SAS 2002
}

%% file: conclusions.tex
\vspace{-1mm}
\section{Conclusions}
\label{sec:conclusions}
\vspace{-1mm}

DSLs provide productivity, portability, and performance by raising the level of
abstraction in the language syntax and semantics, and therefore are a potential
solution to the problem of parallel programming.  However, implementing a high
performance DSL from scratch is not a trivial task, especially when targeting
parallel heterogeneous systems.  To address this issue, we implemented the
Delite Compiler Framework that drastically reduces the effort of building a DSL
by providing an extensible common infrastructure for heterogeneous target code
generation and general/domain-specific optimizations. We presented the
building blocks of the framework and described how they can be easily extended
to build a DSL that runs on heterogeneous hardware.  We demonstrated the
benefits of using the framework with examples from OptiML, a machine learning
DSL implemented with Delite, and showed the performance of OptiML applications
running on a system with multi-core CPUs and a GPU.

\section*{Acknowledgments}
\vspace{-1mm}

The authors would like to thank the DSL'11 reviewers for their high-quality
feedback and Jeremy G.~Siek for shepherding this paper. Thanks also to 
Peter B.~Kessler for reviewing draft versions of this paper.

\vspace{-1mm}